\documentclass[pra,showpacs,showkeys,preprintnumbers,amsmath,amssymb,superscriptaddress,twocolumn]{revtex4-1}
\usepackage[english]{babel}
\usepackage{makeidx} 
\usepackage{graphicx} 
\usepackage{dcolumn} 
\usepackage{array} 
\usepackage{amssymb} 
\usepackage{amsmath}
\usepackage{textcomp}
\usepackage{multirow}
\usepackage{subfigure}
\usepackage{eucal}
\usepackage{mathrsfs}
\usepackage[all]{xy}
\usepackage{epstopdf}

\usepackage{color}

\usepackage{float} 
\usepackage{amsmath} 
\usepackage{amsfonts}
\usepackage{bm}

\begin{document}

\title{Quantifying quantum coherence in a metal-silicate framework}
 
\author{C. Cruz\email{clebson.cruz@ufob.edu.br}} \affiliation{Grupo de Informa\c{c}\~{a}o Qu\^{a}ntica, Centro de Ci\^{e}ncias Exatas e das Tecnologias, Universidade Federal do Oeste da Bahia - Campus Reitor Edgard Santos. Rua Bertioga, 892, Morada Nobre I, 47810-059 Barreiras, Bahia, Brazil.}
\author{M. F. Anka} \affiliation{Instituto de F\'{i}sica, Universidade Federal Fluminense, Av. Gal. Milton Tavares de Souza s/n, 24210-346 Niter\'{o}i, Rio de Janeiro, Brazil.}
\date{\today}
 
\begin{abstract}

The study of quantum coherence in condensed matter systems is a broad avenue to be explored toward the enhancement of its quantum properties by means of material engineering. In this regard, the present work reports a study of the influence of temperature, pressure and magnetic fields on the quantum coherence of a Cu(II) metal-silicate framework. We calculate the $l_1$  trace norm quantum coherence as a function of the magnetic susceptibility of the compound, which allows us to evaluate the effects of these external parameters on the degree of coherence of the material. Our results show that the quantum coherence of a low-dimensional molecular magnetic system can be handled by the management of the external conditions, offering new prospects for quantum coherence measurements through magnetometric experiments. 
\end{abstract}
\maketitle

\section{Introduction}

The development of new technologies has provided great advances in materials preparation techniques, leading to the emergence of new electronic devices. Nowadays, we have reached the point where the miniaturization of these devices has led to the development of molecular size components. However, designing molecular components requires a deep understanding of the quantum properties of these components.
The technological challenges of quantum information science led us to consider fundamental aspects of molecular magnetism, because of its ease of synthesis, great versatility, and low-dimensional quantum features \cite{mario,cruz,cruz2016quantum,diogo,duarte,souza2,mario2,souza,esteves2014new,leite2015heptacopper,Shi2017,cruz2017influence}.

In recent years, it has been demonstrated that molecular magnetic systems present themselves as strong candidates as prototype materials for emerging quantum devices, and the characterization of its quantum correlations has received a considerable attention \cite{cruz,cruz2017influence,diogo,duarte,souza2,mario2,souza,diogo3,duarte2,castro2016thermal}. Recently, it has been demonstrated that this systems may be immune to decoherence mechanisms, presenting highly stable quantum correlations against external perturbations such as temperature and magnetic fields \cite{cruz,cruz2017influence,diogo,duarte,souza2,mario2,souza,diogo3,duarte2,castro2016thermal}. On the other hand, while the entanglement and nonclassical correlations (or quantum correlations) are a key resource to characterize  the quantum properties of a bipartite and some multipartite systems, quantum coherence is a common necessary condition for different forms of quantum correlations \cite{xi2015quantum,hu2018quantum}, being a fundamental feature for signifying quantumness in an integral system \cite{hu2018quantum}.

Quantum coherence, arising from the coherent superposition of quantum states, is a remarkable feature in quantum optics, quantum information theory, solid state physics, quantum game theory, quantum metrology and thermodynamics \cite{PhysRevLett.113.170401,nielsen,giovannetti2011advances,lambert2013quantum,hu2018quantum,theurer2019quantifying,yadin2019coherence,xi2015quantum,streltsov2016quantum,kammerlander2016coherence,goold2016role,santos2020entanglement,passos2019non}. Recently a criterion of measurement that quantifies quantum coherence in solid states system was proposed by Baumgratz \textit{et al.} \cite{baumgratz2014quantifying}. The authors established the fundamental assumptions for a quantitative theory of coherence, enabling the development of a rigorous theory of quantum coherence as a physical resource  \cite{streltsov2016quantum}. However, in order to utilize the remarkable features of quantum coherence it is necessary to define a consistent theoretical basis to measure it experimentally.

In the present work, we report a study of the $l_{1}$ trace norm quantum coherence \cite{streltsov2016quantum,baumgratz2014quantifying,hu2018quantum,rana2016trace} in a antiferromagnetic metal-silicate framework, formed in the compound $KNaCuSi_{4}O_{10}$ \cite{cruz2017influence,brandao2009magnetic}. We establish a relationship between the the measurement of quantum coherence and the magnetic susceptibility of the compound, which allows us to estimate the thermal quantum coherence directly from a set of experimental data. We characterize the degree of quantum coherence in this molecular magnetic system and investigate the influence of the temperature, pressure and magnetic fields. Our results show that it is possible to handle the degree of coherence in a low-dimensional molecular magnetic system by controlling those external conditions, offering a new prospect for the measurement of quantum coherence in condensed matter systems through magnetometric experiments, leading to the development of novel materials with enhanced quantum properties employing materials engineering.

\section{Quantum Coherence in a Molecular Magnetic System}

Geometric approaches are widely used to characterize and quantify the quantum correlations in a wide variety of quantum systems. Similarly to that proposed in the entanglement theory \cite{horodecki}, from which
the entanglement can be characterized by a distance between the considered state and a set of states closed under LOCC operations (separable states) \cite{baumgratz2014quantifying,hu2018quantum,horodecki,vedral1997quantifying,vedral1998entanglement}, Baumgratz et al. \cite{baumgratz2014quantifying} provides one path towards quantifying the amount  of coherence in a quantum state $\rho$. From the minimal distance $D(\rho,\sigma)$, between the considered quantum state $\rho$ and a set $\lbrace \sigma=\sum_{k}^{d} \vert k\rangle\langle k \vert \in \mathcal{I} \rbrace$ of incoherent states of the $d$-dimensional Hilbert space, it is possible to quantify the quantum coherence as:
\begin{eqnarray}
\mathcal{C}_D=\min_{\lbrace \sigma \in \mathcal{I}\rbrace} D(\rho,\sigma)~
\end{eqnarray}
where $D(\rho,\sigma)$ is any measure of distance between the two density matrices, where the reference basis $\lbrace \vert k\rangle \rbrace_{\{k=1,...,d\}}$ may be defined by the physical nature of the problem under investigation or by a task for which coherence is required \cite{streltsov2016quantum,baumgratz2014quantifying}. 

From the selected reference basis, the superposition consists in the nonvanishing off-diagonal terms of the density operator $\rho$ that describes the quantum state of the system of interest \cite{hu2018quantum,baumgratz2014quantifying}. From this consideration Baumgratz et al. \cite{baumgratz2014quantifying}  showed that the $l_{1}$ trace norm can be a reliable measurement of quantum coherence \cite{streltsov2016quantum,baumgratz2014quantifying,hu2018quantum,rana2016trace} as 
\begin{eqnarray}
\mathcal{C}_{l_{1}}&=& \min_{\sigma \in \mathcal{I}} \Vert \rho -\sigma \Vert_{l_1}=\sum_{i\neq j} \vert \langle i\vert \rho\vert j \rangle\vert~.
\label{coherence}
\end{eqnarray}

In order to study the role of external parameters, such as temperature, pressure and magnetic fields, on the quantum coherence of a molecular magnetic system, we will take the reference basis as one of the spin eigenbasis in a certain direction within a quantum metrology setting. 
For the temperature dependence on the quantum coherence, in an experimental point of view, we associate the calculation of the trace norm quantum coherence to the measurement of magnetic susceptibility. 

\subsection{Experimental determination of thermal quantum coherence}

The material, in which we evaluate the quantum coherence, is the KNaCuSi$_{4}$O$_{10}$ compound \cite{hefter1982synthesis,brandao2009magnetic,cruz2017influence}, a metal-silicate framework formed by Cu(II) spin dimers ($d^9$ electronic configuration). This material was chosen because it has synthetic analogs of the naturally occurring mineral litidionite \cite{brandao2009magnetic}, being an ideal realization of a two-qubit system \cite{cruz2017influence}, 
since these dimers are magnetically isolated from each other, separated by two SiO$_4$ corners \cite{brandao2009magnetic}.

Therefore, this prototype material can be described by two-qubit system interacting by the Heisenberg model \cite{mario,sarandy,brandao2009magnetic,cruz2017influence}, 
\begin{equation}
    \mathcal{H}=-J \vec{S}_1\cdot \vec{S}_2~,
    \label{hamiltoniana}
\end{equation}
where $J$ is the coupling constant.
The magnetic susceptibility of this prototype material corresponds to the Bleaney-Bowers equation \cite{mario,bleaney}:
\begin{equation}
\chi (T)=\frac{2 N(g\mu_B)^2}{k_B T}\frac{1}{3+e^{-{J}/{k_B T}}}~,
\label{eq:4}
\end{equation}
where $g$ is the Land\'{e} factor, $\mu_B$ is the Bohr magneton, $k_B$ is the Boltzmann constant and $N$ is the number of dimers.
Magnetic susceptibility measurements for the KNaCuSi$_{4}$O$_{10}$ compound are presented in reference \onlinecite{brandao2009magnetic}, where the measurements have been performed between 2 K and 350 K. Taking into account the crystal structure of this compound, the susceptibility data have been fitted to the Bleany-Bowers equation, Eq. (\ref{eq:4}), where the authors obtain $J/k_B=-2.86(3)$ K (antiferromagnetic coupled ions). 

To go further and analyze the quantum coherence by means of the magnetic susceptibility of the material, firstly, we write the density matrix of the system under consideration in the local $S_z$ eigenbasis $\lbrace \vert 00\rangle,\vert 01\rangle,\vert 10\rangle,\vert 11\rangle\rbrace$, \cite{yuri,yuri2,cruz2017influence} :
\begin{eqnarray}
\rho(T)&=&\frac{1}{4}\left[
   \begin{matrix} 1+c(T) &   &   &   \\
                        & 1-c(T) & 2c(T) &   \\
                        & 2c(T) & 1-c(T) &   \\
                        &   &   & 1+c(T)
   \end{matrix} \right]
   \label{eq:1}
\end{eqnarray}
where,
\begin{equation}
c(T)=\frac{2k_BT\chi(T)}{N_Ag^2\mu_B^2}-1
\label{eq:3}
\end{equation} 
is the correlation function between the spins \cite{cruz,yuri,yuri2,cruz2017influence}. The isotropy of the magnetic material and the rotation symmetry of the Heisenberg interaction, Eq. (\ref{hamiltoniana}), at zero field makes the density matrix, Eq. (\ref{eq:1}), the same Bell's diagonal mixed state for any local $S^{(x)}$, $S^{(y)}$ or $S^{(z)}$ eigenbasis. Thus, for the isotropic Heisenberg interaction, the coherence will be basis independent for any of these spin eigenbasis.

From Eq. (\ref{coherence}) and (\ref{eq:1}) one can evaluate the temperature dependence on the $l_{1}$ norm quantum coherence as a function of the temperature, measured by the magnetic susceptibility of the compound as 
\begin{eqnarray}
\mathcal{C}(T)&=& \left| \frac{2k_BT\chi(T)}{N_Ag^2\mu_B^2}-1\right|~.
\label{csus}
\end{eqnarray}
Therefore, from Eq. (\ref{csus}), it is possible to measure quantum coherence in a low dimensional molecular magnetic system by measuring the thermodynamic properties of solids, such as magnetic susceptibility. 

\begin{figure}[!h]
\centering
{\includegraphics[scale=0.3]{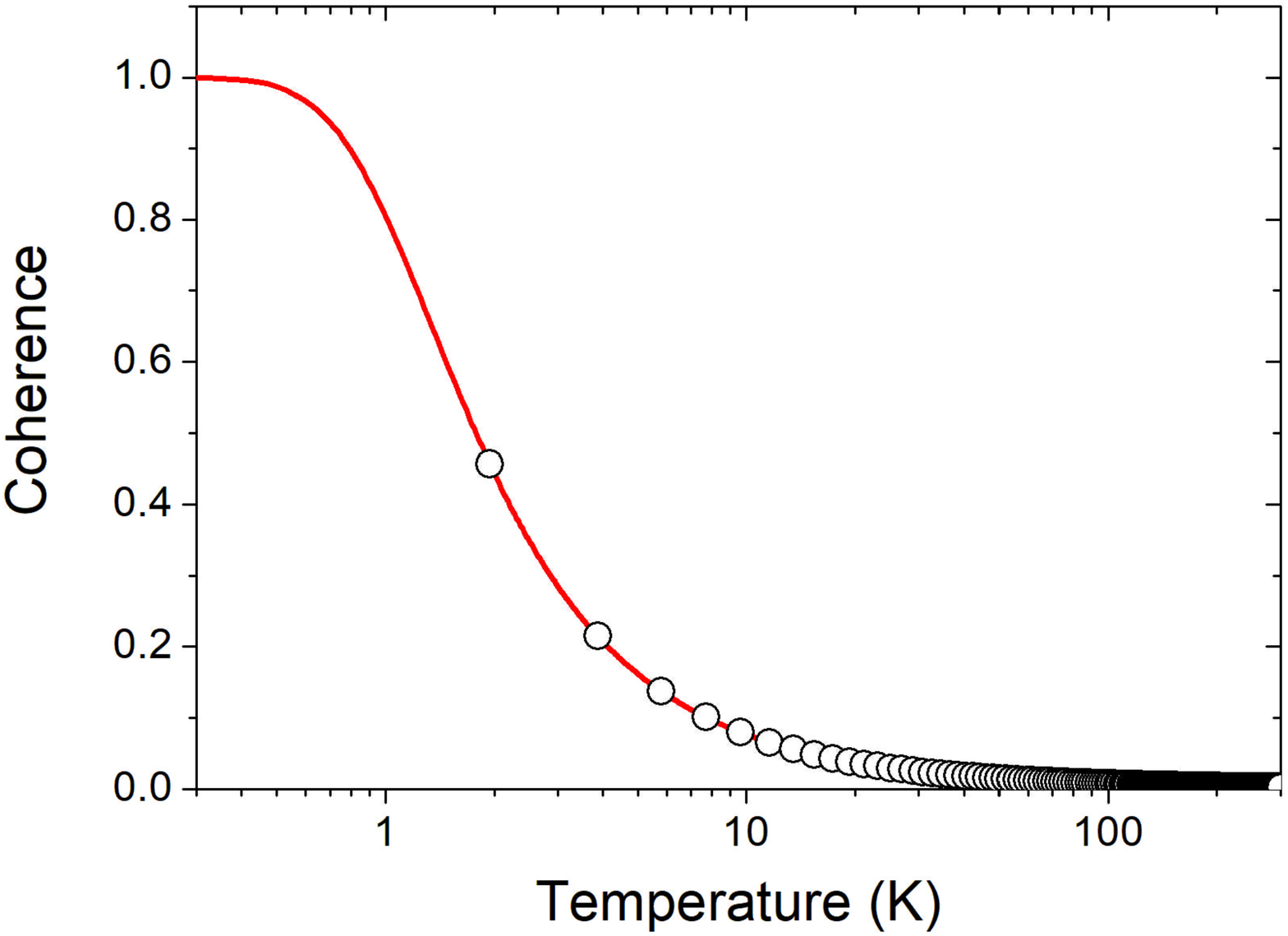}}\\
\caption{Experimental (open circles) and theoretical (solid red line) quantum coherence as a function of the temperature, measured by the magnetic susceptibility of the KNaCuSi$_{4}$O$_{10}$ metal-silicate framework \cite{brandao2009magnetic}.}
\label{fig2}
\end{figure}

In Fig. \ref{fig2}, we show the quantum coherence obtained from the measurement of the magnetic susceptibility of our prototype material KNaCuSi$_{4}$O$_{10}$, reported in reference \onlinecite{brandao2009magnetic}. The theoretical curves were plotted taking the corresponding estimates for the coupling constant $J/k_B$ in Eq. (\ref{csus}). The reference \onlinecite{diogo} investigates the thermal quantum entanglement of this prototype material finding the maximum temperature of $2.43(7)$ K below which there is quantum entanglement between the Cu(II) ions; at this temperature the coherence of the system is $35\%$ of the maximum value. As can be seen, as expected, increasing the temperature changes the Boltzmann's weights due to the thermal fluctuations, which changes the occupation of the energy levels, leading the system to populate incoherent states.

On the other hand, apart from its quantification and characterization, the coherence is intimately associated with other quantum correlations quantifiers \cite{ma2016converting,adesso2016measures}. Thus, quantum coherence and correlations can be transformed into each other. The quantum discord \cite{zurek}, a measurement of quantum correlations beyond entanglement \cite{streltsov2016quantum,cruz,adesso2016measures}, can also be quantified in molecular magnetic system via a distance based approach \cite{cruz,cruz2017influence,cruz2016quantum}. In this regard, Eq. \ref{csus} relates to the geometric quantum discord based on the Schatten 1-norm \cite{ciccarello} as
\begin{eqnarray}
\mathcal{Q}_G(\rho)=\min_{\omega_{cq}}\Vert\rho - \rho_c\Vert\label{eq:11}= \frac{\mathcal{C}(T)}{2}~,
\end{eqnarray}
where $\Vert A\Vert_1=\mbox{Tr}\left[\sqrt{A^\dagger A}\right]$, $\rho$ is a given quantum state and $\rho_c$ is the closest classical-quantum state \cite{sarandy2,sarandy3}. Therefore, the measurement of quantum coherence can also quantify the amount of quantum correlation in a molecular magnetic system.

\subsection{Influence of the external pressure}

Recently, using Density Funtional Theory calculations \cite{hohenberg1964inhomogeneous}, one of us shown that the application of an hydrostatic pressure in the prototype material KNaCuSi$_{4}$O$_{10}$ induces a structural contraction, which leads to a minimization on the degree of its quantum correlations \cite{cruz2017influence}. Through the dependence of the magnetic coupling constant of the compound with the external pressure \cite{cruz2017influence} we calculate magnetic susceptibilities, Eq. (\ref{eq:4}), from each magnetic coupling constant. Thus, by using Eqs. (\ref{csus}) it is possible to evaluate the influence of an hydrostatic pressure on the quantum coherence of a low dimensional molecular magnetic system. In this way, we establish a relationship between the quantum coherence and significant macroscopic effects, as an external hydrostatic pressure applied on the magnetic material.

\begin{figure}[!h]
\centering
\includegraphics[scale=0.335]{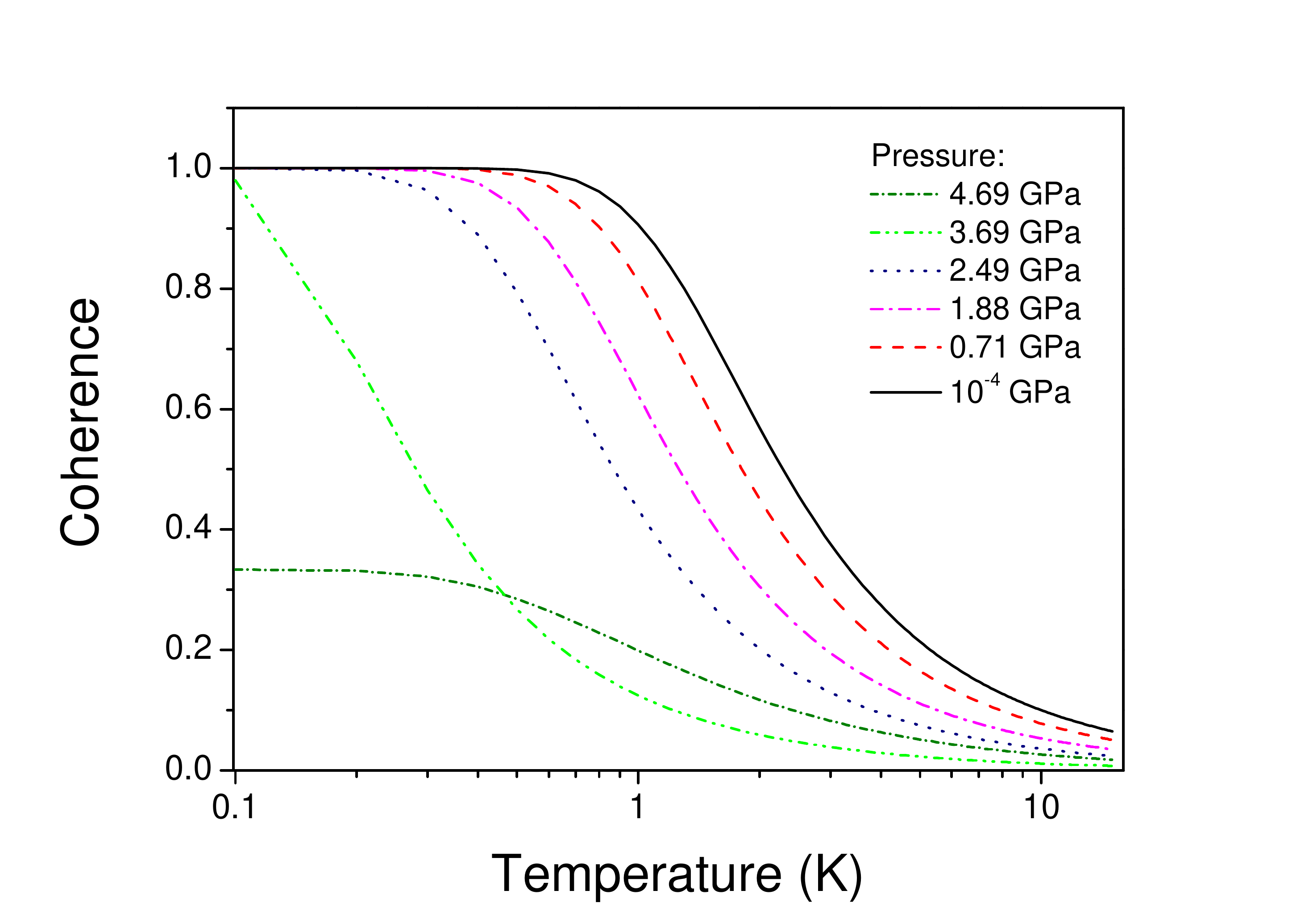}\\
\caption{Temperature dependence of the quantum coherence calculated for different values of hydrostatic pressure. This result was obtained through the dependence of the magnetic coupling constant $J$ with the external pressure, which was obtained by Density Funtional Theory calculations the KNaCuSi$_{4}$O$_{10}$ metal-silicate framework, reported in reference \onlinecite{cruz2017influence}. Solid black line is the coherence ontained at ambient pressure shown in figure \ref{fig2}.}
\label{fig:pressure}
\end{figure}

Figure \ref{fig:pressure} shows temperature dependence on the quantum coherence for different values of hydrostatic pressure.
As obtained in the reference \onlinecite{cruz2017influence}, increasing the pressure in the system leads to a decrease on the lattice parameter and volume of unit cell; as a consequence, the exchange parameter of the system increases until it becomes positive changing the energy levels, i.e., the system ceases to be ordered antiferromagnetically in an entangled ground state $\left[\vert 01 \rangle - \vert 10 \rangle\right]/\sqrt{2}$ and is led to populate ferromagnetically ordered states with a lower degree of coherence. Due to this change of the exchange parameter sign, there is a gap in the coherence of the ground state, as it can be seen in figure \ref{fig:pressure}, increasing the hydrostatic pressure on this prototype material leads to the decrease of the degree of coherence in the system.

Therefore, increasing the pressure changes the magnetic alignment, which yields a change on the occupation of the energy levels, leading the system to a less coherent configuration. 
On this matter, it is possible to handle the degree of coherence of a low dimensional molecular magnetic system by managing the pressure applied on the system, since the external pressure induces a structural contraction in the metal--silicate framework, leading to a change of its magnetic alignment and reducing the degree of quantum coherence in the dimeric unit. This result shows that the degree of coherence in a spin cluster system can be controlled by the management of structural parameters such as the lattice parameter and volume of the unit cell, allowing the manipulation of the quantum coherence by materials engineering.

\subsection{Influence of the longitudinal and transverse magnetic field}

Unlike other quantum information theoretic quantifiers, quantum coherence is basis dependent. The reference basis $\lbrace{\vert k \rangle}\rbrace$ concerning the coherence is measured in regard to the physical problem under investigation; e.g., for molecular magnetic systems the usual basis is the spin eigenbasis  in a certain direction, $\lbrace S_x,S_y,S_z\rbrace$, within a quantum metrology setting.

In order to investigate the influence of the application of an external magnetic field on the prototype material KNaCuSi$_{4}$O$_{10}$, we consider the magnetic field $\vec{B}$ along the $z$ direction. One can evaluate the coherence for two reference bases: one parallel to the applied field ($S_z$ eigenbasis) and another perpendicular to the field ($S_x$ or $S_y$ eigenbasis). 
The Hamiltonian that rules this system interacting with an external magnetic field is given by
\begin{eqnarray}
\mathcal{H} = -J\vec{S_1}\cdot\vec{S_2} - \mu_B g \vec{B}\cdot \left( \vec{S}_1 + \vec{S}_1 \right) ~.
\label{model}
\end{eqnarray}

Thus, our aim is study how the applied fields, parallel to the density matrix basis $S_z$ (longitudinal field) and perpendicular to the density matrix basis $S_x$ (transverse field), affect the degree of quantum coherence in our prototype material.

\subsubsection{Longitudinal Field}

The bipartite density matrix of this system can be written in the $S^{(z)}$ eigenbasis: $\lbrace \vert 00\rangle,\vert 01\rangle,\vert 10\rangle,\vert 11\rangle\rbrace$ \cite{yuri,yuri2} as a X-shaped matrix:
\begin{eqnarray}
\rho_z(T,B_z) &=&\frac{e^{x}}{2Z}\left[
   \begin{matrix} 2e^{\beta h_z} & & &  \\
                        & {1 +  e^{-4x}} & {1 -  e^{-4x}}  & \\
                        & {1 -  e^{-4x}}  & {1 +  e^{-4x}}  & \\\
                        & & & 2e^{-\beta h_z} 
   \end{matrix} \right]~.
\label{longitudinal}
\end{eqnarray}
where
\begin{equation}
    Z(T,B_z)= e^{x} + e^{-3x} + 2 e^{x} \cosh\left(\beta h_z\right)~,
\end{equation}
with $\beta = 1/k_BT$, $x=\beta J/4$, $h_z = \mu_Bg_zB_z$, and $Z$
is the partition function. 

The trace norm quantum coherence, Eq.(\ref{coherence}), can be calculated in terms of these matrix elements as:
\begin{eqnarray}
\mathcal{C}_z(T,B_z)&=& \left|\frac{1 -  e^{-4x}}{1 + e^{-4x} + 2 \cosh\left(\beta h_z\right)}\right|~.
\label{cz}
\end{eqnarray}

\begin{figure}[!h]
\centering
\subfigure[]{\includegraphics[scale=0.495]{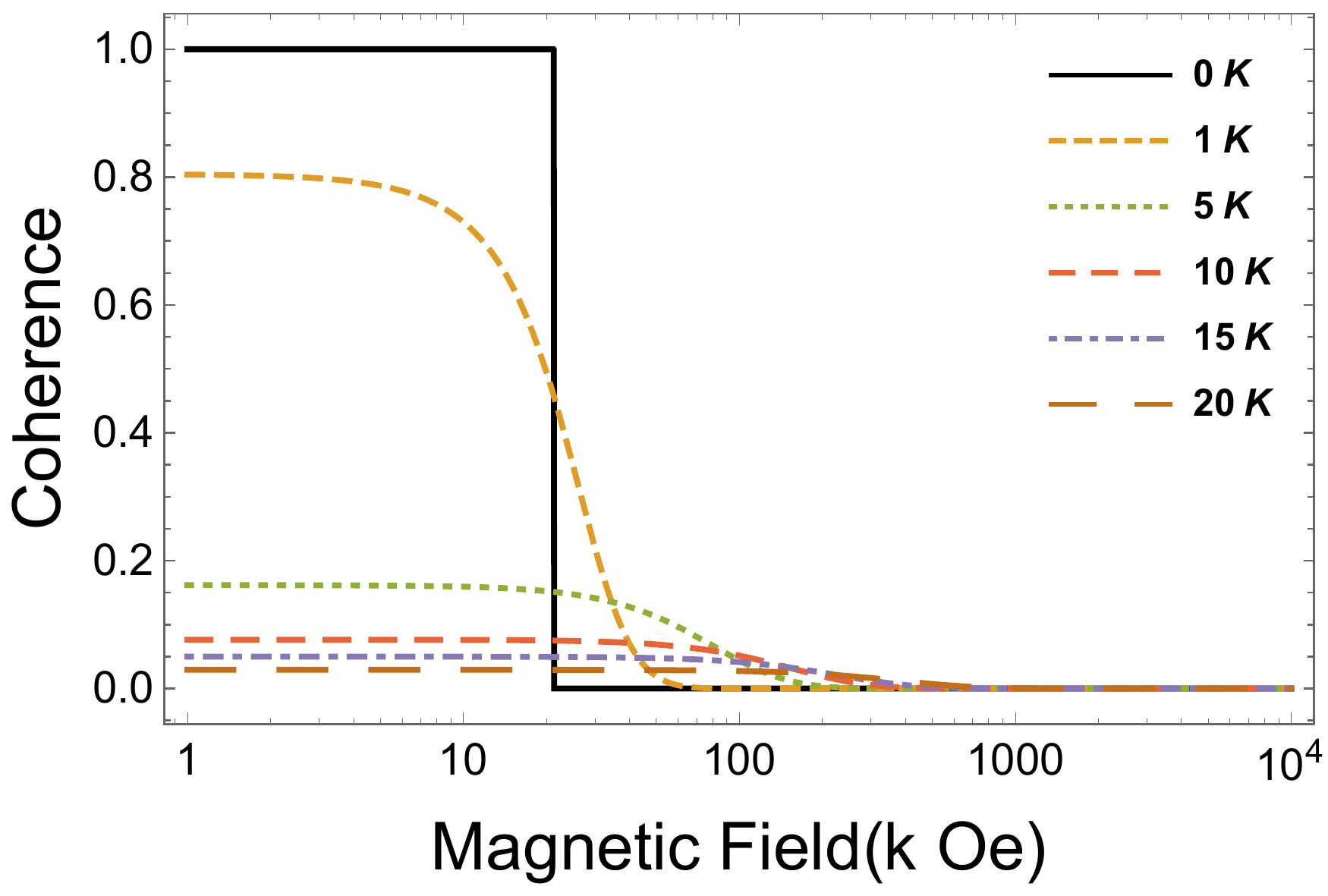}}\\
\subfigure[]{\includegraphics[scale=0.49]{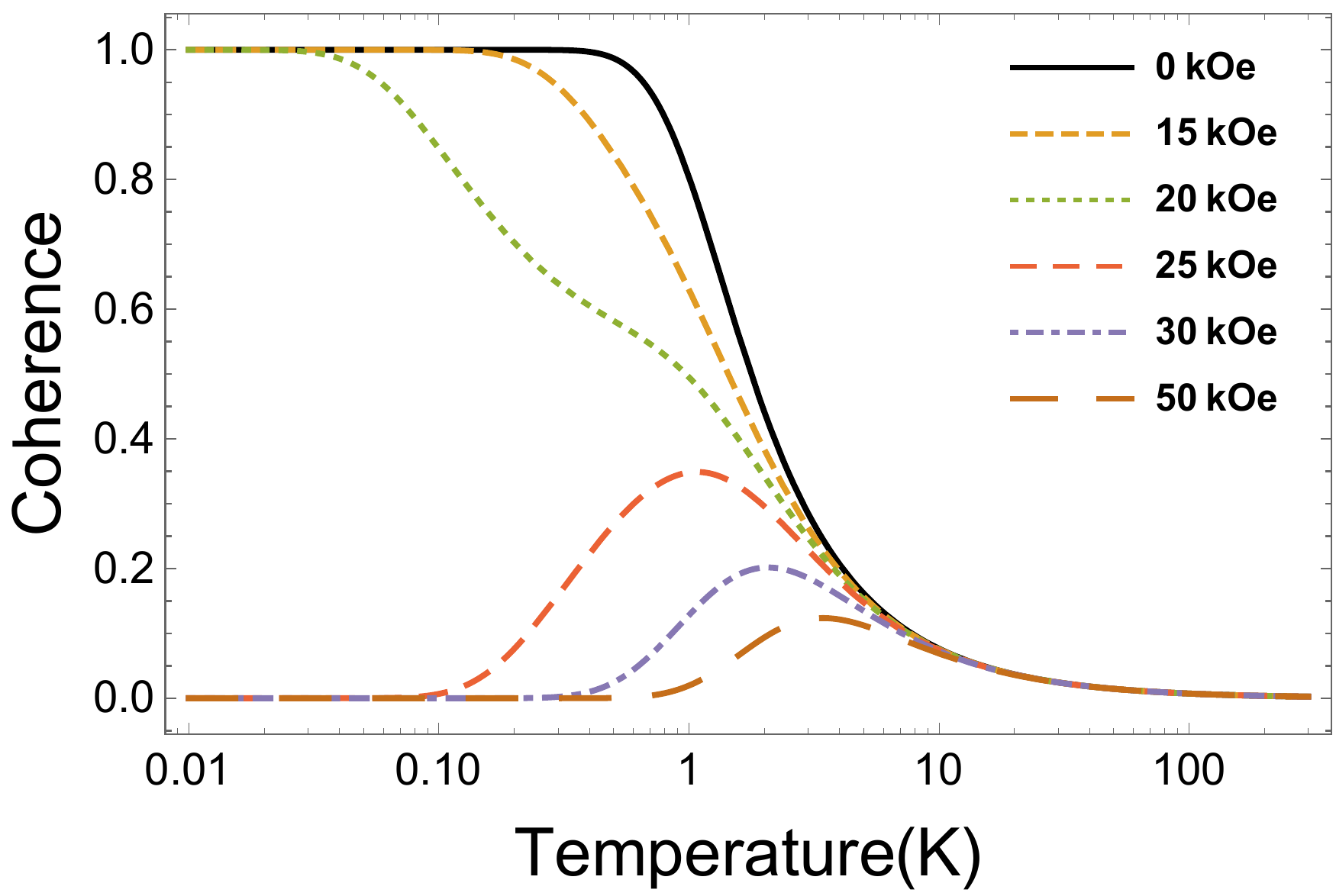}}\\
\caption{(Color online) Quantum coherence for selected temperatures as a function of the applied field in the (a) parallel eigenbasis $S_z$ (longitudinal field) and (b) the perpendicular eigenbasis $S_x$ (transverse field).}
\label{figlong}
\end{figure}

Figure \ref{figlong} shows the quantum coherence of our prototype material, as a function of the applied longitudinal field (figure \ref{figlong}(a)) and the temperature (figure \ref{figlong}(b)). As can be seen, the application of longitudinal magnetic field decreases the degree of coherence. As a system in equilibrium at 0 K is always in its maximally entangled ground state $\left[\vert 01 \rangle - \vert 10 \rangle\right]/\sqrt{2}$, since it is an antiferromagnetic $1/2$-spin dimer \cite{brandao2009magnetic,cruz2017influence}, when the longitudinal field reaches a critical value of $B_c=21 279$ Oe the system will be led to the incoherent ground state $\vert 00 \rangle$, i.e, all spins aligned with the applied field. Hence, there is an abrupt change in the degree of coherence of the ground state, the system changes from a maximally entangled ground state with $\mathcal{C}_z=1$  ($B < B_c$) to an completely incoherent ground state ($B \geq B_c$), due to the Zeeman effect along the parallel direction with the density matrix basis. Therefore, the aplication of a longitudinal magnetic field changes the energy eigenvalues, which leads the system to a different ground state with a smaller degree of coherence. 

Although absolute zero is not physically realizable, the changes on the degree of coherence between the Cu(II) ions, due to the application of the longitudinal, are highlighted in low temperatures. At nonzero temperatures, the thermal fluctuations compete with the magnetic interactions changing 
the occupation of the energy levels \cite{souza2,maziero2010quantum,castro2016thermal}. Therefore, increasing the temperature will populate many incoherent states, decreasing the degree of coherence of the system. In contrast, when a high longitudinal magnetic field is applied at low temperatures the ground state tends to be less coherent than some excited states. Thus, increasing the temperature may lead the system to some excited states with a larger degree of coherence, yeldind a small increase on the quantum coherence, as can be seen in figure \ref{figlong}(b) around 1 K. Similar effects were also encountered for the entanglement of formation and quantum discord in the references \onlinecite{souza2,maziero2010quantum,arnesen2001natural,asoudeh2006thermal}.

\subsubsection{Transverse Field}

On the other hand, the field applied on $z$ direction breaks the isotropy of the Heisenberg interaction due to the Zeeman effect along this direction \cite{mario}. In this regard, in order to study the effects of the application of a transverse magnetic field on the quantum coherence of KNaCuSi$_{4}$O$_{10}$ compound, we rewrite the density matrix  Eq.(\ref{longitudinal}) in the perpendicular eigenbasis, $S^{(x)}$, $\lbrace \vert ++\rangle,\vert +-\rangle,\vert -+\rangle,\vert --\rangle\rbrace$ as:
\begin{widetext}
\begin{eqnarray}
\rho_x(T,B_z) = \dfrac{e^{x}}{2Z} \left[\begin{matrix}
 \cosh(\beta h_z) + 1 & \sinh(\beta h_z) & \sinh(\beta h_z) & \cosh(\beta h_z) - 1 \\
\sinh(\beta h_z) &\cosh(\beta h_z) + e^{-4x} & \cosh(\beta h_z) - e^{-4x} & \sinh(\beta h_z) \\
\sinh(\beta h_z) & \cosh(\beta h_z) - e^{-4x} & \cosh(\beta h_z) + e^{-4x} & \sinh(\beta h_z)\\
\cosh(\beta h_z) - 1 & \sinh(\beta h_z) & \sinh(\beta h_z) & \cosh(\beta h_z) + 1
\end{matrix}\right]~.
\label{transverse}
\end{eqnarray}
\end{widetext}

From Eq. (\ref{transverse}) one can found that the transverse field quantum coherence as:
\begin{eqnarray}
\mathcal{C}_x (T,B_z) &=& \frac{e^{x}}{Z}\left(\left|\cosh(\beta h_z)-1\right| +4 \left|\sinh(\beta h_z)\right|+ \right. \nonumber \\ 
&& \left. + \left|\cosh(\beta h_z) - e^{-4x}\right|\right)~. \label{transversez}
\end{eqnarray}

\begin{figure}[!h]
\centering
\subfigure[]{\includegraphics[scale=0.577]{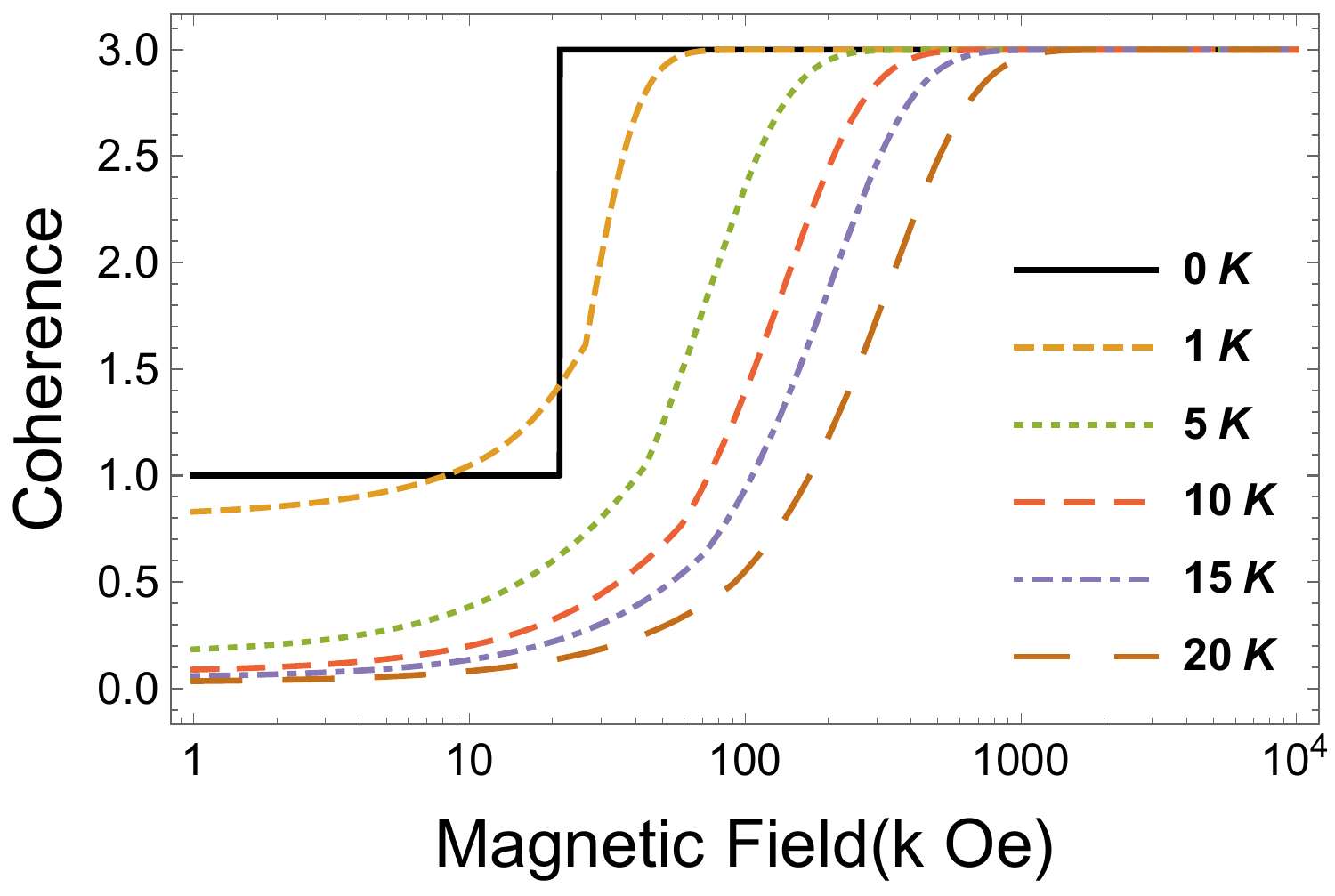}}\\
\subfigure[]{\includegraphics[scale=0.567]{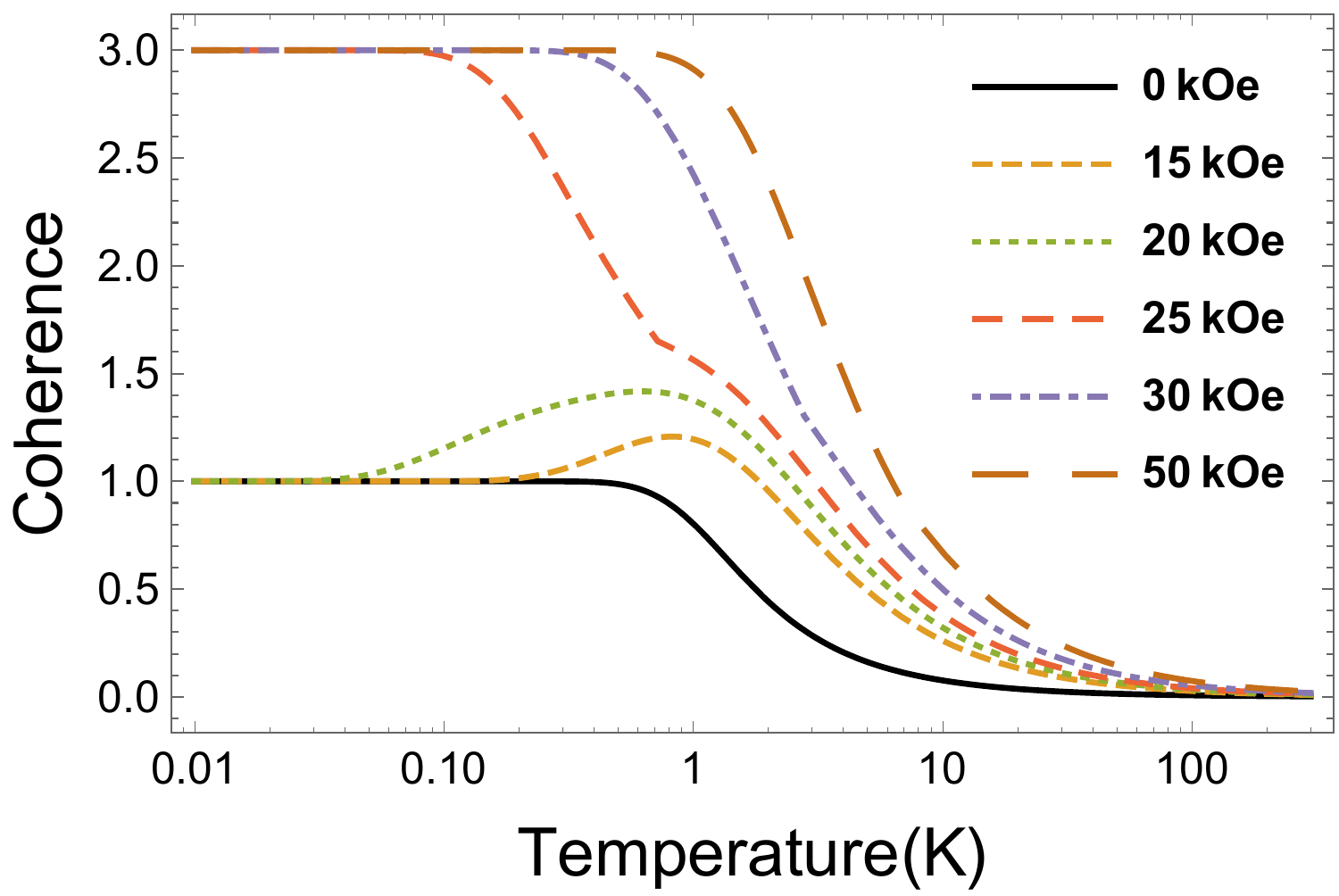}}\\
\caption{(Color online) Temperature dependence of quantum coherence in the (a) parallel eigenbasis $S_z$ (longitudinal field) and (b) the perpendicular eigenbasis $S_x$ (transverse field), for selected magnetic fields.}
\label{figtrans}
\end{figure}

In Figure \ref{figtrans}, we show the quantum coherence of our metal-silicate framework as a function of the applied transverse field (figure \ref{figtrans}(a)) and the temperature (figure \ref{figtrans}(b)). In contrast to the application of a longitudinal magnetic field (figure \ref{figlong}), the  transverse field increases and strengthens the degree of coherence between Cu(II) ions in the dimeric cluster. It also can be understood in terms of the population change of the ground state, due to the variation of the Boltzmann's weights, which leads to a change on the occupation of the energy levels \cite{souza2,maziero2010quantum,castro2016thermal}. 

Due to the rotational invariance of the Bell states, at 0 K and $B_z < B_c$the system is found in the the maximally entangled ground state  $\left[\vert +- \rangle - \vert -+ \rangle\right]/\sqrt{2}$ on the $S^{(x)}$ eigenbasis, with quantum coherence $\mathcal{C}_z=1$.
For the transverse field $B_z \geq B_c$, the system will be led to the ground state $\vert 00 \rangle = \left[\vert ++ \rangle + \vert +- \rangle + \vert -+ \rangle + \vert -- \rangle\right]/2$, which has the maximal coherence degree $\mathcal{C}_z=3$, i.e., the ground state becomes maximally coherent beyond the coherence of the maximally entangled ground state. Thus, this metal-silicate framework changes from a maximally entangled ground state $\mathcal{C}_z=1$ ($B < B_c$) to a maximally coherent ground state with $\mathcal{C}_z=3$ ($B \geq B_c$) even at nonzero temperatures (figure \ref{figtrans}(b)). This transition, induced by the application of a transverse magnetic field, describes an abrupt change in the degree of coherence of the ground state, due to the uncertainty relations between the  $S_z$ spin direction of the Zeeman effect and the $S_x$ eigenbasis, in which the density matrix is written, Eq. (\ref{transverse}). This yields a change in the populations of the energy levels, leading the system to a state with the highest degree of coherence. Therefore, the transverse field effect is to populate such coherent states, leading to an increase in the degree coherence of the system.

It is worth noting that, for $T>10$ K the values of magnetic field needed to increase or decrease the coherence in this metal-silicate framework, are higher than the available field intensities usually created in laboratories, which indicates that the coherence in this low-dimensional molecular magnetic system cannot be destroyed  or enhanced by a common longitudinal or transverse magnetic field, respectively, for temperatures above 10 K, providing the quantum coherence of this low-dimensional molecular magnetic system is resistant to magnetic field application above this temperature.

\section{Conclusions}

In summary, we have shown a method to evaluate the degree of quantum coherence in low dimensional magnetic materials composed of 1/2-spin dimers, where we investigate the influence of external parameters, such as temperature and magnetic fields, on the quantum coherence of the $KNaCuSi_{4}O_{10}$ prototype material, which is an ideal simulation of a two-qubit system. At zero magnetic field we associated the calculation of the trace norm quantum coherence with  a magnetometric measurement of the magnetic susceptibility of the compound, using the  experimental data reported in reference \onlinecite{brandao2009magnetic}, which characterizes the magnetic properties of this material. We established the theoretical relations between the measurement of quantum coherence and the thermodynamic properties of this prototype material, setting the path for the experimental measurement of quantum coherence in low dimensional molecular magnetic materials. 

In addition, we investigate the influence of the applied pressure on the degree of quantum coherence of our prototype material. We observed that increasing the external pressure yields a reduction on the degree of coherence in the system due to structural contraction provided by applying external pressure. Therefore, the quantum coherence of a molecular magnetic system can be handled by the management structural properties of the material.
Moreover, we also presented a theoretical analysis of the influence of an external magnetic field on the degree of coherence of the system, where we investigate the basis dependence on the coherence. We found that when the longitudinally field is applied to the reference basis, it decreases the degree of coherence in the parallel eigenbasis, wereas if the transversely field is applied it strengthens the degree of coherence between the Cu(II) ions in the dimeric cluster in the transverse eigenbasis.

In this context, the coherence of a low-dimensional molecular magnetic system can be handled by controlling the thermodynamic's external parameters, such as temperature, external pressure and magnetic fields. These results offer a new prospect for quantum coherence measurement, leading to promising applications in quantum information science such as the enhancement of quantum properties in low-dimensional molecular magnetic systems by materials engineering, and the development of novel candidate platforms for processing and transmission of quantum information. 

\begin{acknowledgments}
The author would like to thank D. O. Soares-Pinto for his helpful comments, and A. M. dos Santos and M. S. Reis for the magnetic susceptibility measurements. This study was financed in part by the National Counsel of Technological and Scientific Development (CNPq) and  the \textit{Coordena\c{c}\~{a}o de Aperfei\c{c}oamento de Pessoal de N\'{i}vel Superior - Brasil} (CAPES) - Finance Code 001.
\end{acknowledgments}


\begin{thebibliography}{49}%
\makeatletter
\providecommand \@ifxundefined [1]{%
 \@ifx{#1\undefined}
}%
\providecommand \@ifnum [1]{%
 \ifnum #1\expandafter \@firstoftwo
 \else \expandafter \@secondoftwo
 \fi
}%
\providecommand \@ifx [1]{%
 \ifx #1\expandafter \@firstoftwo
 \else \expandafter \@secondoftwo
 \fi
}%
\providecommand \natexlab [1]{#1}%
\providecommand \enquote  [1]{``#1''}%
\providecommand \bibnamefont  [1]{#1}%
\providecommand \bibfnamefont [1]{#1}%
\providecommand \citenamefont [1]{#1}%
\providecommand \href@noop [0]{\@secondoftwo}%
\providecommand \href [0]{\begingroup \@sanitize@url \@href}%
\providecommand \@href[1]{\@@startlink{#1}\@@href}%
\providecommand \@@href[1]{\endgroup#1\@@endlink}%
\providecommand \@sanitize@url [0]{\catcode `\\12\catcode `\$12\catcode
  `\&12\catcode `\#12\catcode `\^12\catcode `\_12\catcode `\%12\relax}%
\providecommand \@@startlink[1]{}%
\providecommand \@@endlink[0]{}%
\providecommand \url  [0]{\begingroup\@sanitize@url \@url }%
\providecommand \@url [1]{\endgroup\@href {#1}{\urlprefix }}%
\providecommand \urlprefix  [0]{URL }%
\providecommand \Eprint [0]{\href }%
\providecommand \doibase [0]{http://dx.doi.org/}%
\providecommand \selectlanguage [0]{\@gobble}%
\providecommand \bibinfo  [0]{\@secondoftwo}%
\providecommand \bibfield  [0]{\@secondoftwo}%
\providecommand \translation [1]{[#1]}%
\providecommand \BibitemOpen [0]{}%
\providecommand \bibitemStop [0]{}%
\providecommand \bibitemNoStop [0]{.\EOS\space}%
\providecommand \EOS [0]{\spacefactor3000\relax}%
\providecommand \BibitemShut  [1]{\csname bibitem#1\endcsname}%
\let\auto@bib@innerbib\@empty
\bibitem [{\citenamefont {Reis}(2013)}]{mario}%
  \BibitemOpen
  \bibfield  {author} {\bibinfo {author} {\bibfnamefont {M.}~\bibnamefont
  {Reis}},\ }\href@noop {} {\emph {\bibinfo {title} {Fundamentals of
  magnetism}}}\ (\bibinfo  {publisher} {Elsevier},\ \bibinfo {year}
  {2013})\BibitemShut {NoStop}%
\bibitem [{\citenamefont {Cruz}\ \emph {et~al.}(2016)\citenamefont {Cruz},
  \citenamefont {Soares-Pinto}, \citenamefont {Brand�o}, \citenamefont {dos
  Santos},\ and\ \citenamefont {Reis}}]{cruz}%
  \BibitemOpen
  \bibfield  {author} {\bibinfo {author} {\bibfnamefont {C.}~\bibnamefont
  {Cruz}}, \bibinfo {author} {\bibfnamefont {D.~O.}\ \bibnamefont
  {Soares-Pinto}}, \bibinfo {author} {\bibfnamefont {P.}~\bibnamefont
  {Brand�o}}, \bibinfo {author} {\bibfnamefont {A.~M.}\ \bibnamefont {dos
  Santos}}, \ and\ \bibinfo {author} {\bibfnamefont {M.~S.}\ \bibnamefont
  {Reis}},\ }\href@noop {} {\bibfield  {journal} {\bibinfo  {journal} {EPL
  (Europhysics Letters)}\ }\textbf {\bibinfo {volume} {113}},\ \bibinfo {pages}
  {40004} (\bibinfo {year} {2016})}\BibitemShut {NoStop}%
\bibitem [{\citenamefont {Cruz}(2017)}]{cruz2016quantum}%
  \BibitemOpen
  \bibfield  {author} {\bibinfo {author} {\bibfnamefont {C.}~\bibnamefont
  {Cruz}},\ }\href {\doibase 10.1142/S0219749917500319} {\bibfield  {journal}
  {\bibinfo  {journal} {International Journal of Quantum Information}\ ,\
  \bibinfo {pages} {1750031}} (\bibinfo {year} {2017})}\BibitemShut {NoStop}%
\bibitem [{\citenamefont {Soares-Pinto}\ \emph {et~al.}(2009)\citenamefont
  {Soares-Pinto}, \citenamefont {Souza}, \citenamefont {Sarthour},
  \citenamefont {Oliveira}, \citenamefont {Reis}, \citenamefont {Brandao},
  \citenamefont {Rocha},\ and\ \citenamefont {dos Santos}}]{diogo}%
  \BibitemOpen
  \bibfield  {author} {\bibinfo {author} {\bibfnamefont {D.}~\bibnamefont
  {Soares-Pinto}}, \bibinfo {author} {\bibfnamefont {A.}~\bibnamefont {Souza}},
  \bibinfo {author} {\bibfnamefont {R.}~\bibnamefont {Sarthour}}, \bibinfo
  {author} {\bibfnamefont {I.}~\bibnamefont {Oliveira}}, \bibinfo {author}
  {\bibfnamefont {M.}~\bibnamefont {Reis}}, \bibinfo {author} {\bibfnamefont
  {P.}~\bibnamefont {Brandao}}, \bibinfo {author} {\bibfnamefont
  {J.}~\bibnamefont {Rocha}}, \ and\ \bibinfo {author} {\bibfnamefont
  {A.}~\bibnamefont {dos Santos}},\ }\href@noop {} {\bibfield  {journal}
  {\bibinfo  {journal} {EPL (Europhysics Letters)}\ }\textbf {\bibinfo {volume}
  {87}},\ \bibinfo {pages} {40008} (\bibinfo {year} {2009})}\BibitemShut
  {NoStop}%
\bibitem [{\citenamefont {Duarte}\ \emph
  {et~al.}(2013{\natexlab{a}})\citenamefont {Duarte}, \citenamefont {Castro},
  \citenamefont {Soares-Pinto},\ and\ \citenamefont {Reis}}]{duarte}%
  \BibitemOpen
  \bibfield  {author} {\bibinfo {author} {\bibfnamefont {O.~S.}\ \bibnamefont
  {Duarte}}, \bibinfo {author} {\bibfnamefont {C.~S.}\ \bibnamefont {Castro}},
  \bibinfo {author} {\bibfnamefont {D.~O.}\ \bibnamefont {Soares-Pinto}}, \
  and\ \bibinfo {author} {\bibfnamefont {M.~S.}\ \bibnamefont {Reis}},\ }\href
  {http://stacks.iop.org/0295-5075/103/i=4/a=40002} {\bibfield  {journal}
  {\bibinfo  {journal} {EPL (Europhysics Letters)}\ }\textbf {\bibinfo {volume}
  {103}},\ \bibinfo {pages} {40002} (\bibinfo {year}
  {2013}{\natexlab{a}})}\BibitemShut {NoStop}%
\bibitem [{\citenamefont {Souza}\ \emph {et~al.}(2008)\citenamefont {Souza},
  \citenamefont {Reis}, \citenamefont {Soares-Pinto}, \citenamefont
  {Oliveira},\ and\ \citenamefont {Sarthour}}]{souza2}%
  \BibitemOpen
  \bibfield  {author} {\bibinfo {author} {\bibfnamefont {A.~M.}\ \bibnamefont
  {Souza}}, \bibinfo {author} {\bibfnamefont {M.~S.}\ \bibnamefont {Reis}},
  \bibinfo {author} {\bibfnamefont {D.~O.}\ \bibnamefont {Soares-Pinto}},
  \bibinfo {author} {\bibfnamefont {I.~S.}\ \bibnamefont {Oliveira}}, \ and\
  \bibinfo {author} {\bibfnamefont {R.~S.}\ \bibnamefont {Sarthour}},\
  }\href@noop {} {\bibfield  {journal} {\bibinfo  {journal} {Physical Review
  B}\ }\textbf {\bibinfo {volume} {77}},\ \bibinfo {pages} {104402} (\bibinfo
  {year} {2008})}\BibitemShut {NoStop}%
\bibitem [{\citenamefont {Reis}\ \emph {et~al.}(2012)\citenamefont {Reis},
  \citenamefont {Soriano}, \citenamefont {dos Santos}, \citenamefont {Sales},
  \citenamefont {Soares-Pinto},\ and\ \citenamefont {Brandao}}]{mario2}%
  \BibitemOpen
  \bibfield  {author} {\bibinfo {author} {\bibfnamefont {M.~S.}\ \bibnamefont
  {Reis}}, \bibinfo {author} {\bibfnamefont {S.}~\bibnamefont {Soriano}},
  \bibinfo {author} {\bibfnamefont {A.~M.}\ \bibnamefont {dos Santos}},
  \bibinfo {author} {\bibfnamefont {B.~C.}\ \bibnamefont {Sales}}, \bibinfo
  {author} {\bibfnamefont {D.}~\bibnamefont {Soares-Pinto}}, \ and\ \bibinfo
  {author} {\bibfnamefont {P.}~\bibnamefont {Brandao}},\ }\href@noop {}
  {\bibfield  {journal} {\bibinfo  {journal} {EPL (Europhysics Letters)}\
  }\textbf {\bibinfo {volume} {100}},\ \bibinfo {pages} {50001} (\bibinfo
  {year} {2012})}\BibitemShut {NoStop}%
\bibitem [{\citenamefont {Souza}\ \emph {et~al.}(2009)\citenamefont {Souza},
  \citenamefont {Soares-Pinto}, \citenamefont {Sarthour}, \citenamefont
  {Oliveira}, \citenamefont {Reis}, \citenamefont {Brandao},\ and\
  \citenamefont {dos Santos}}]{souza}%
  \BibitemOpen
  \bibfield  {author} {\bibinfo {author} {\bibfnamefont {A.~M.}\ \bibnamefont
  {Souza}}, \bibinfo {author} {\bibfnamefont {D.~O.}\ \bibnamefont
  {Soares-Pinto}}, \bibinfo {author} {\bibfnamefont {R.~S.}\ \bibnamefont
  {Sarthour}}, \bibinfo {author} {\bibfnamefont {I.~S.}\ \bibnamefont
  {Oliveira}}, \bibinfo {author} {\bibfnamefont {M.~S.}\ \bibnamefont {Reis}},
  \bibinfo {author} {\bibfnamefont {P.}~\bibnamefont {Brandao}}, \ and\
  \bibinfo {author} {\bibfnamefont {A.~M.}\ \bibnamefont {dos Santos}},\
  }\href@noop {} {\bibfield  {journal} {\bibinfo  {journal} {Physical Review
  B}\ }\textbf {\bibinfo {volume} {79}},\ \bibinfo {pages} {054408} (\bibinfo
  {year} {2009})}\BibitemShut {NoStop}%
\bibitem [{\citenamefont {Esteves}\ \emph {et~al.}(2014)\citenamefont
  {Esteves}, \citenamefont {Tedesco}, \citenamefont {Pedro}, \citenamefont
  {Cruz}, \citenamefont {Reis},\ and\ \citenamefont
  {Brandao}}]{esteves2014new}%
  \BibitemOpen
  \bibfield  {author} {\bibinfo {author} {\bibfnamefont {D.}~\bibnamefont
  {Esteves}}, \bibinfo {author} {\bibfnamefont {J.}~\bibnamefont {Tedesco}},
  \bibinfo {author} {\bibfnamefont {S.}~\bibnamefont {Pedro}}, \bibinfo
  {author} {\bibfnamefont {C.}~\bibnamefont {Cruz}}, \bibinfo {author}
  {\bibfnamefont {M.}~\bibnamefont {Reis}}, \ and\ \bibinfo {author}
  {\bibfnamefont {P.}~\bibnamefont {Brandao}},\ }\href@noop {} {\bibfield
  {journal} {\bibinfo  {journal} {Materials Chemistry and Physics}\ }\textbf
  {\bibinfo {volume} {147}},\ \bibinfo {pages} {611} (\bibinfo {year}
  {2014})}\BibitemShut {NoStop}%
\bibitem [{\citenamefont {Leite~Ferreira}\ \emph {et~al.}(2015)\citenamefont
  {Leite~Ferreira}, \citenamefont {Brand{\~a}o}, \citenamefont {Dos~Santos},
  \citenamefont {Gai}, \citenamefont {Cruz}, \citenamefont {Reis},
  \citenamefont {Santos},\ and\ \citenamefont
  {F{\'e}lix}}]{leite2015heptacopper}%
  \BibitemOpen
  \bibfield  {author} {\bibinfo {author} {\bibfnamefont {B.}~\bibnamefont
  {Leite~Ferreira}}, \bibinfo {author} {\bibfnamefont {P.}~\bibnamefont
  {Brand{\~a}o}}, \bibinfo {author} {\bibfnamefont {A.}~\bibnamefont
  {Dos~Santos}}, \bibinfo {author} {\bibfnamefont {Z.}~\bibnamefont {Gai}},
  \bibinfo {author} {\bibfnamefont {C.}~\bibnamefont {Cruz}}, \bibinfo {author}
  {\bibfnamefont {M.}~\bibnamefont {Reis}}, \bibinfo {author} {\bibfnamefont
  {T.}~\bibnamefont {Santos}}, \ and\ \bibinfo {author} {\bibfnamefont
  {V.}~\bibnamefont {F{\'e}lix}},\ }\href@noop {} {\bibfield  {journal}
  {\bibinfo  {journal} {Journal of Coordination Chemistry}\ }\textbf {\bibinfo
  {volume} {68}},\ \bibinfo {pages} {2770} (\bibinfo {year}
  {2015})}\BibitemShut {NoStop}%
\bibitem [{\citenamefont {Shi}\ \emph {et~al.}(2017)\citenamefont {Shi},
  \citenamefont {Bai}, \citenamefont {Lu}, \citenamefont {Cruz}, \citenamefont
  {Reis},\ and\ \citenamefont {Gao}}]{Shi2017}%
  \BibitemOpen
  \bibfield  {author} {\bibinfo {author} {\bibfnamefont {F.-N.}\ \bibnamefont
  {Shi}}, \bibinfo {author} {\bibfnamefont {Y.-W.}\ \bibnamefont {Bai}},
  \bibinfo {author} {\bibfnamefont {M.}~\bibnamefont {Lu}}, \bibinfo {author}
  {\bibfnamefont {C.}~\bibnamefont {Cruz}}, \bibinfo {author} {\bibfnamefont
  {M.~S.}\ \bibnamefont {Reis}}, \ and\ \bibinfo {author} {\bibfnamefont
  {J.}~\bibnamefont {Gao}},\ }\href {\doibase 10.1007/s11243-017-0165-5}
  {\bibfield  {journal} {\bibinfo  {journal} {Transition Metal Chemistry}\ }
  (\bibinfo {year} {2017}),\ 10.1007/s11243-017-0165-5}\BibitemShut {NoStop}%
\bibitem [{\citenamefont {Cruz}\ \emph {et~al.}(2017)\citenamefont {Cruz},
  \citenamefont {Alves}, \citenamefont {dos Santos}, \citenamefont
  {Soares-Pinto}, \citenamefont {de~Jesus}, \citenamefont {de~Almeida},\ and\
  \citenamefont {Reis}}]{cruz2017influence}%
  \BibitemOpen
  \bibfield  {author} {\bibinfo {author} {\bibfnamefont {C.}~\bibnamefont
  {Cruz}}, \bibinfo {author} {\bibfnamefont {{\'A}.}~\bibnamefont {Alves}},
  \bibinfo {author} {\bibfnamefont {R.}~\bibnamefont {dos Santos}}, \bibinfo
  {author} {\bibfnamefont {D.}~\bibnamefont {Soares-Pinto}}, \bibinfo {author}
  {\bibfnamefont {J.}~\bibnamefont {de~Jesus}}, \bibinfo {author}
  {\bibfnamefont {J.}~\bibnamefont {de~Almeida}}, \ and\ \bibinfo {author}
  {\bibfnamefont {M.}~\bibnamefont {Reis}},\ }\href@noop {} {\bibfield
  {journal} {\bibinfo  {journal} {EPL (Europhysics Letters)}\ }\textbf
  {\bibinfo {volume} {117}},\ \bibinfo {pages} {20004} (\bibinfo {year}
  {2017})}\BibitemShut {NoStop}%
\bibitem [{\citenamefont {Soares-Pinto}\ \emph {et~al.}(2011)\citenamefont
  {Soares-Pinto}, \citenamefont {Teles}, \citenamefont {Souza}, \citenamefont
  {Deazevedo}, \citenamefont {Sarthour}, \citenamefont {Bonagamba},
  \citenamefont {Reis},\ and\ \citenamefont {Oliveira}}]{diogo3}%
  \BibitemOpen
  \bibfield  {author} {\bibinfo {author} {\bibfnamefont {D.}~\bibnamefont
  {Soares-Pinto}}, \bibinfo {author} {\bibfnamefont {J.}~\bibnamefont {Teles}},
  \bibinfo {author} {\bibfnamefont {A.}~\bibnamefont {Souza}}, \bibinfo
  {author} {\bibfnamefont {E.}~\bibnamefont {Deazevedo}}, \bibinfo {author}
  {\bibfnamefont {R.}~\bibnamefont {Sarthour}}, \bibinfo {author}
  {\bibfnamefont {T.}~\bibnamefont {Bonagamba}}, \bibinfo {author}
  {\bibfnamefont {M.}~\bibnamefont {Reis}}, \ and\ \bibinfo {author}
  {\bibfnamefont {I.}~\bibnamefont {Oliveira}},\ }\href@noop {} {\bibfield
  {journal} {\bibinfo  {journal} {International Journal of Quantum
  Information}\ }\textbf {\bibinfo {volume} {9}},\ \bibinfo {pages} {1047}
  (\bibinfo {year} {2011})}\BibitemShut {NoStop}%
\bibitem [{\citenamefont {Duarte}\ \emph
  {et~al.}(2013{\natexlab{b}})\citenamefont {Duarte}, \citenamefont {Castro},\
  and\ \citenamefont {Reis}}]{duarte2}%
  \BibitemOpen
  \bibfield  {author} {\bibinfo {author} {\bibfnamefont {O.~S.}\ \bibnamefont
  {Duarte}}, \bibinfo {author} {\bibfnamefont {C.~S.}\ \bibnamefont {Castro}},
  \ and\ \bibinfo {author} {\bibfnamefont {M.~S.}\ \bibnamefont {Reis}},\
  }\href@noop {} {\bibfield  {journal} {\bibinfo  {journal} {Physical Review
  A}\ }\textbf {\bibinfo {volume} {88}},\ \bibinfo {pages} {012317} (\bibinfo
  {year} {2013}{\natexlab{b}})}\BibitemShut {NoStop}%
\bibitem [{\citenamefont {Castro}\ \emph {et~al.}(2016)\citenamefont {Castro},
  \citenamefont {Duarte}, \citenamefont {Pires}, \citenamefont {Soares-Pinto},\
  and\ \citenamefont {Reis}}]{castro2016thermal}%
  \BibitemOpen
  \bibfield  {author} {\bibinfo {author} {\bibfnamefont {C.}~\bibnamefont
  {Castro}}, \bibinfo {author} {\bibfnamefont {O.}~\bibnamefont {Duarte}},
  \bibinfo {author} {\bibfnamefont {D.}~\bibnamefont {Pires}}, \bibinfo
  {author} {\bibfnamefont {D.}~\bibnamefont {Soares-Pinto}}, \ and\ \bibinfo
  {author} {\bibfnamefont {M.}~\bibnamefont {Reis}},\ }\href@noop {} {\bibfield
   {journal} {\bibinfo  {journal} {Physics Letters A}\ }\textbf {\bibinfo
  {volume} {380}},\ \bibinfo {pages} {1571} (\bibinfo {year}
  {2016})}\BibitemShut {NoStop}%
\bibitem [{\citenamefont {Xi}\ \emph {et~al.}(2015)\citenamefont {Xi},
  \citenamefont {Li},\ and\ \citenamefont {Fan}}]{xi2015quantum}%
  \BibitemOpen
  \bibfield  {author} {\bibinfo {author} {\bibfnamefont {Z.}~\bibnamefont
  {Xi}}, \bibinfo {author} {\bibfnamefont {Y.}~\bibnamefont {Li}}, \ and\
  \bibinfo {author} {\bibfnamefont {H.}~\bibnamefont {Fan}},\ }\href@noop {}
  {\bibfield  {journal} {\bibinfo  {journal} {Scientific reports}\ }\textbf
  {\bibinfo {volume} {5}},\ \bibinfo {pages} {10922} (\bibinfo {year}
  {2015})}\BibitemShut {NoStop}%
\bibitem [{\citenamefont {Hu}\ \emph {et~al.}(2018)\citenamefont {Hu},
  \citenamefont {Hu}, \citenamefont {Wang}, \citenamefont {Peng}, \citenamefont
  {Zhang},\ and\ \citenamefont {Fan}}]{hu2018quantum}%
  \BibitemOpen
  \bibfield  {author} {\bibinfo {author} {\bibfnamefont {M.-L.}\ \bibnamefont
  {Hu}}, \bibinfo {author} {\bibfnamefont {X.}~\bibnamefont {Hu}}, \bibinfo
  {author} {\bibfnamefont {J.}~\bibnamefont {Wang}}, \bibinfo {author}
  {\bibfnamefont {Y.}~\bibnamefont {Peng}}, \bibinfo {author} {\bibfnamefont
  {Y.-R.}\ \bibnamefont {Zhang}}, \ and\ \bibinfo {author} {\bibfnamefont
  {H.}~\bibnamefont {Fan}},\ }\href@noop {} {\bibfield  {journal} {\bibinfo
  {journal} {Physics Reports}\ } (\bibinfo {year} {2018})}\BibitemShut
  {NoStop}%
\bibitem [{\citenamefont {Girolami}(2014)}]{PhysRevLett.113.170401}%
  \BibitemOpen
  \bibfield  {author} {\bibinfo {author} {\bibfnamefont {D.}~\bibnamefont
  {Girolami}},\ }\href {\doibase 10.1103/PhysRevLett.113.170401} {\bibfield
  {journal} {\bibinfo  {journal} {Phys. Rev. Lett.}\ }\textbf {\bibinfo
  {volume} {113}},\ \bibinfo {pages} {170401} (\bibinfo {year}
  {2014})}\BibitemShut {NoStop}%
\bibitem [{\citenamefont {Nielsen}\ and\ \citenamefont
  {Chuang}(2010)}]{nielsen}%
  \BibitemOpen
  \bibfield  {author} {\bibinfo {author} {\bibfnamefont {M.~A.}\ \bibnamefont
  {Nielsen}}\ and\ \bibinfo {author} {\bibfnamefont {I.~L.}\ \bibnamefont
  {Chuang}},\ }\href@noop {} {\emph {\bibinfo {title} {Quantum computation and
  quantum information}}}\ (\bibinfo  {publisher} {Cambridge university press},\
  \bibinfo {year} {2010})\BibitemShut {NoStop}%
\bibitem [{\citenamefont {Giovannetti}\ \emph {et~al.}(2011)\citenamefont
  {Giovannetti}, \citenamefont {Lloyd},\ and\ \citenamefont
  {Maccone}}]{giovannetti2011advances}%
  \BibitemOpen
  \bibfield  {author} {\bibinfo {author} {\bibfnamefont {V.}~\bibnamefont
  {Giovannetti}}, \bibinfo {author} {\bibfnamefont {S.}~\bibnamefont {Lloyd}},
  \ and\ \bibinfo {author} {\bibfnamefont {L.}~\bibnamefont {Maccone}},\
  }\href@noop {} {\bibfield  {journal} {\bibinfo  {journal} {Nature photonics}\
  }\textbf {\bibinfo {volume} {5}},\ \bibinfo {pages} {222} (\bibinfo {year}
  {2011})}\BibitemShut {NoStop}%
\bibitem [{\citenamefont {Lambert}\ \emph {et~al.}(2013)\citenamefont
  {Lambert}, \citenamefont {Chen}, \citenamefont {Cheng}, \citenamefont {Li},
  \citenamefont {Chen},\ and\ \citenamefont {Nori}}]{lambert2013quantum}%
  \BibitemOpen
  \bibfield  {author} {\bibinfo {author} {\bibfnamefont {N.}~\bibnamefont
  {Lambert}}, \bibinfo {author} {\bibfnamefont {Y.-N.}\ \bibnamefont {Chen}},
  \bibinfo {author} {\bibfnamefont {Y.-C.}\ \bibnamefont {Cheng}}, \bibinfo
  {author} {\bibfnamefont {C.-M.}\ \bibnamefont {Li}}, \bibinfo {author}
  {\bibfnamefont {G.-Y.}\ \bibnamefont {Chen}}, \ and\ \bibinfo {author}
  {\bibfnamefont {F.}~\bibnamefont {Nori}},\ }\href@noop {} {\bibfield
  {journal} {\bibinfo  {journal} {Nature Physics}\ }\textbf {\bibinfo {volume}
  {9}},\ \bibinfo {pages} {10} (\bibinfo {year} {2013})}\BibitemShut {NoStop}%
\bibitem [{\citenamefont {Theurer}\ \emph {et~al.}(2019)\citenamefont
  {Theurer}, \citenamefont {Egloff}, \citenamefont {Zhang},\ and\ \citenamefont
  {Plenio}}]{theurer2019quantifying}%
  \BibitemOpen
  \bibfield  {author} {\bibinfo {author} {\bibfnamefont {T.}~\bibnamefont
  {Theurer}}, \bibinfo {author} {\bibfnamefont {D.}~\bibnamefont {Egloff}},
  \bibinfo {author} {\bibfnamefont {L.}~\bibnamefont {Zhang}}, \ and\ \bibinfo
  {author} {\bibfnamefont {M.~B.}\ \bibnamefont {Plenio}},\ }\href@noop {}
  {\bibfield  {journal} {\bibinfo  {journal} {Physical Review Letters}\
  }\textbf {\bibinfo {volume} {122}},\ \bibinfo {pages} {190405} (\bibinfo
  {year} {2019})}\BibitemShut {NoStop}%
\bibitem [{\citenamefont {Yadin}\ \emph {et~al.}(2019)\citenamefont {Yadin},
  \citenamefont {Bogaert}, \citenamefont {Susa},\ and\ \citenamefont
  {Girolami}}]{yadin2019coherence}%
  \BibitemOpen
  \bibfield  {author} {\bibinfo {author} {\bibfnamefont {B.}~\bibnamefont
  {Yadin}}, \bibinfo {author} {\bibfnamefont {P.}~\bibnamefont {Bogaert}},
  \bibinfo {author} {\bibfnamefont {C.~E.}\ \bibnamefont {Susa}}, \ and\
  \bibinfo {author} {\bibfnamefont {D.}~\bibnamefont {Girolami}},\ }\href@noop
  {} {\bibfield  {journal} {\bibinfo  {journal} {Physical Review A}\ }\textbf
  {\bibinfo {volume} {99}},\ \bibinfo {pages} {012329} (\bibinfo {year}
  {2019})}\BibitemShut {NoStop}%
\bibitem [{\citenamefont {Streltsov}\ \emph {et~al.}(2017)\citenamefont
  {Streltsov}, \citenamefont {Adesso},\ and\ \citenamefont
  {Plenio}}]{streltsov2016quantum}%
  \BibitemOpen
  \bibfield  {author} {\bibinfo {author} {\bibfnamefont {A.}~\bibnamefont
  {Streltsov}}, \bibinfo {author} {\bibfnamefont {G.}~\bibnamefont {Adesso}}, \
  and\ \bibinfo {author} {\bibfnamefont {M.~B.}\ \bibnamefont {Plenio}},\
  }\href@noop {} {\bibfield  {journal} {\bibinfo  {journal} {Reviews of Modern
  Physics}\ }\textbf {\bibinfo {volume} {89}},\ \bibinfo {pages} {041003}
  (\bibinfo {year} {2017})}\BibitemShut {NoStop}%
\bibitem [{\citenamefont {Kammerlander}\ and\ \citenamefont
  {Anders}(2016)}]{kammerlander2016coherence}%
  \BibitemOpen
  \bibfield  {author} {\bibinfo {author} {\bibfnamefont {P.}~\bibnamefont
  {Kammerlander}}\ and\ \bibinfo {author} {\bibfnamefont {J.}~\bibnamefont
  {Anders}},\ }\href@noop {} {\bibfield  {journal} {\bibinfo  {journal}
  {Scientific reports}\ }\textbf {\bibinfo {volume} {6}},\ \bibinfo {pages}
  {22174} (\bibinfo {year} {2016})}\BibitemShut {NoStop}%
\bibitem [{\citenamefont {Goold}\ \emph {et~al.}(2016)\citenamefont {Goold},
  \citenamefont {Huber}, \citenamefont {Riera}, \citenamefont {del Rio},\ and\
  \citenamefont {Skrzypczyk}}]{goold2016role}%
  \BibitemOpen
  \bibfield  {author} {\bibinfo {author} {\bibfnamefont {J.}~\bibnamefont
  {Goold}}, \bibinfo {author} {\bibfnamefont {M.}~\bibnamefont {Huber}},
  \bibinfo {author} {\bibfnamefont {A.}~\bibnamefont {Riera}}, \bibinfo
  {author} {\bibfnamefont {L.}~\bibnamefont {del Rio}}, \ and\ \bibinfo
  {author} {\bibfnamefont {P.}~\bibnamefont {Skrzypczyk}},\ }\href@noop {}
  {\bibfield  {journal} {\bibinfo  {journal} {Journal of Physics A:
  Mathematical and Theoretical}\ }\textbf {\bibinfo {volume} {49}},\ \bibinfo
  {pages} {143001} (\bibinfo {year} {2016})}\BibitemShut {NoStop}%
\bibitem [{\citenamefont {Santos}(2020)}]{santos2020entanglement}%
  \BibitemOpen
  \bibfield  {author} {\bibinfo {author} {\bibfnamefont {A.~C.}\ \bibnamefont
  {Santos}},\ }\href@noop {} {\bibfield  {journal} {\bibinfo  {journal}
  {Quantum Information Processing}\ }\textbf {\bibinfo {volume} {19}},\
  \bibinfo {pages} {13} (\bibinfo {year} {2020})}\BibitemShut {NoStop}%
\bibitem [{\citenamefont {Passos}\ \emph {et~al.}(2019)\citenamefont {Passos},
  \citenamefont {Obando}, \citenamefont {Balthazar}, \citenamefont {Paula},
  \citenamefont {Huguenin},\ and\ \citenamefont {Sarandy}}]{passos2019non}%
  \BibitemOpen
  \bibfield  {author} {\bibinfo {author} {\bibfnamefont {M.}~\bibnamefont
  {Passos}}, \bibinfo {author} {\bibfnamefont {P.~C.}\ \bibnamefont {Obando}},
  \bibinfo {author} {\bibfnamefont {W.}~\bibnamefont {Balthazar}}, \bibinfo
  {author} {\bibfnamefont {F.}~\bibnamefont {Paula}}, \bibinfo {author}
  {\bibfnamefont {J.}~\bibnamefont {Huguenin}}, \ and\ \bibinfo {author}
  {\bibfnamefont {M.}~\bibnamefont {Sarandy}},\ }\href@noop {} {\bibfield
  {journal} {\bibinfo  {journal} {Optics letters}\ }\textbf {\bibinfo {volume}
  {44}},\ \bibinfo {pages} {2478} (\bibinfo {year} {2019})}\BibitemShut
  {NoStop}%
\bibitem [{\citenamefont {Baumgratz}\ \emph {et~al.}(2014)\citenamefont
  {Baumgratz}, \citenamefont {Cramer},\ and\ \citenamefont
  {Plenio}}]{baumgratz2014quantifying}%
  \BibitemOpen
  \bibfield  {author} {\bibinfo {author} {\bibfnamefont {T.}~\bibnamefont
  {Baumgratz}}, \bibinfo {author} {\bibfnamefont {M.}~\bibnamefont {Cramer}}, \
  and\ \bibinfo {author} {\bibfnamefont {M.~B.}\ \bibnamefont {Plenio}},\
  }\href@noop {} {\bibfield  {journal} {\bibinfo  {journal} {Physical review
  letters}\ }\textbf {\bibinfo {volume} {113}},\ \bibinfo {pages} {140401}
  (\bibinfo {year} {2014})}\BibitemShut {NoStop}%
\bibitem [{\citenamefont {Rana}\ \emph {et~al.}(2016)\citenamefont {Rana},
  \citenamefont {Parashar},\ and\ \citenamefont {Lewenstein}}]{rana2016trace}%
  \BibitemOpen
  \bibfield  {author} {\bibinfo {author} {\bibfnamefont {S.}~\bibnamefont
  {Rana}}, \bibinfo {author} {\bibfnamefont {P.}~\bibnamefont {Parashar}}, \
  and\ \bibinfo {author} {\bibfnamefont {M.}~\bibnamefont {Lewenstein}},\
  }\href@noop {} {\bibfield  {journal} {\bibinfo  {journal} {Physical Review
  A}\ }\textbf {\bibinfo {volume} {93}},\ \bibinfo {pages} {012110} (\bibinfo
  {year} {2016})}\BibitemShut {NoStop}%
\bibitem [{\citenamefont {Brandao}\ \emph {et~al.}(2009)\citenamefont
  {Brandao}, \citenamefont {Rocha}, \citenamefont {Reis}, \citenamefont
  {Dos~Santos},\ and\ \citenamefont {Jin}}]{brandao2009magnetic}%
  \BibitemOpen
  \bibfield  {author} {\bibinfo {author} {\bibfnamefont {P.}~\bibnamefont
  {Brandao}}, \bibinfo {author} {\bibfnamefont {J.}~\bibnamefont {Rocha}},
  \bibinfo {author} {\bibfnamefont {M.~S.}\ \bibnamefont {Reis}}, \bibinfo
  {author} {\bibfnamefont {A.}~\bibnamefont {Dos~Santos}}, \ and\ \bibinfo
  {author} {\bibfnamefont {R.}~\bibnamefont {Jin}},\ }\href@noop {} {\bibfield
  {journal} {\bibinfo  {journal} {Journal of Solid State Chemistry}\ }\textbf
  {\bibinfo {volume} {182}},\ \bibinfo {pages} {253} (\bibinfo {year}
  {2009})}\BibitemShut {NoStop}%
\bibitem [{\citenamefont {Horodecki}\ \emph {et~al.}(2009)\citenamefont
  {Horodecki}, \citenamefont {Horodecki}, \citenamefont {Horodecki},\ and\
  \citenamefont {Horodecki}}]{horodecki}%
  \BibitemOpen
  \bibfield  {author} {\bibinfo {author} {\bibfnamefont {R.}~\bibnamefont
  {Horodecki}}, \bibinfo {author} {\bibfnamefont {P.}~\bibnamefont
  {Horodecki}}, \bibinfo {author} {\bibfnamefont {M.}~\bibnamefont
  {Horodecki}}, \ and\ \bibinfo {author} {\bibfnamefont {K.}~\bibnamefont
  {Horodecki}},\ }\href@noop {} {\bibfield  {journal} {\bibinfo  {journal}
  {Reviews of modern physics}\ }\textbf {\bibinfo {volume} {81}},\ \bibinfo
  {pages} {865} (\bibinfo {year} {2009})}\BibitemShut {NoStop}%
\bibitem [{\citenamefont {Vedral}\ \emph {et~al.}(1997)\citenamefont {Vedral},
  \citenamefont {Plenio}, \citenamefont {Rippin},\ and\ \citenamefont
  {Knight}}]{vedral1997quantifying}%
  \BibitemOpen
  \bibfield  {author} {\bibinfo {author} {\bibfnamefont {V.}~\bibnamefont
  {Vedral}}, \bibinfo {author} {\bibfnamefont {M.~B.}\ \bibnamefont {Plenio}},
  \bibinfo {author} {\bibfnamefont {M.~A.}\ \bibnamefont {Rippin}}, \ and\
  \bibinfo {author} {\bibfnamefont {P.~L.}\ \bibnamefont {Knight}},\
  }\href@noop {} {\bibfield  {journal} {\bibinfo  {journal} {Physical Review
  Letters}\ }\textbf {\bibinfo {volume} {78}},\ \bibinfo {pages} {2275}
  (\bibinfo {year} {1997})}\BibitemShut {NoStop}%
\bibitem [{\citenamefont {Vedral}\ and\ \citenamefont
  {Plenio}(1998)}]{vedral1998entanglement}%
  \BibitemOpen
  \bibfield  {author} {\bibinfo {author} {\bibfnamefont {V.}~\bibnamefont
  {Vedral}}\ and\ \bibinfo {author} {\bibfnamefont {M.~B.}\ \bibnamefont
  {Plenio}},\ }\href@noop {} {\bibfield  {journal} {\bibinfo  {journal}
  {Physical Review A}\ }\textbf {\bibinfo {volume} {57}},\ \bibinfo {pages}
  {1619} (\bibinfo {year} {1998})}\BibitemShut {NoStop}%
\bibitem [{\citenamefont {Hefter}\ and\ \citenamefont
  {Kenney}(1982)}]{hefter1982synthesis}%
  \BibitemOpen
  \bibfield  {author} {\bibinfo {author} {\bibfnamefont {J.}~\bibnamefont
  {Hefter}}\ and\ \bibinfo {author} {\bibfnamefont {M.~E.}\ \bibnamefont
  {Kenney}},\ }\href@noop {} {\bibfield  {journal} {\bibinfo  {journal}
  {Inorganic Chemistry}\ }\textbf {\bibinfo {volume} {21}},\ \bibinfo {pages}
  {2810} (\bibinfo {year} {1982})}\BibitemShut {NoStop}%
\bibitem [{\citenamefont {Sarandy}(2009)}]{sarandy}%
  \BibitemOpen
  \bibfield  {author} {\bibinfo {author} {\bibfnamefont {M.~S.}\ \bibnamefont
  {Sarandy}},\ }\href@noop {} {\bibfield  {journal} {\bibinfo  {journal}
  {Physical Review A}\ }\textbf {\bibinfo {volume} {80}},\ \bibinfo {pages}
  {022108} (\bibinfo {year} {2009})}\BibitemShut {NoStop}%
\bibitem [{\citenamefont {Bleaney}\ and\ \citenamefont
  {Bowers}(1952)}]{bleaney}%
  \BibitemOpen
  \bibfield  {author} {\bibinfo {author} {\bibfnamefont {B.}~\bibnamefont
  {Bleaney}}\ and\ \bibinfo {author} {\bibfnamefont {K.}~\bibnamefont
  {Bowers}},\ }in\ \href@noop {} {\emph {\bibinfo {booktitle} {Proceedings of
  the Royal Society of London A: Mathematical, Physical and Engineering
  Sciences}}},\ Vol.\ \bibinfo {volume} {214}\ (\bibinfo {organization} {The
  Royal Society},\ \bibinfo {year} {1952})\ pp.\ \bibinfo {pages}
  {451--465}\BibitemShut {NoStop}%
\bibitem [{\citenamefont {Yurishchev}(2011)}]{yuri}%
  \BibitemOpen
  \bibfield  {author} {\bibinfo {author} {\bibfnamefont {M.~A.}\ \bibnamefont
  {Yurishchev}},\ }\href@noop {} {\bibfield  {journal} {\bibinfo  {journal}
  {Physical Review B}\ }\textbf {\bibinfo {volume} {84}},\ \bibinfo {pages}
  {024418} (\bibinfo {year} {2011})}\BibitemShut {NoStop}%
\bibitem [{\citenamefont {Aldoshin}\ \emph {et~al.}(2014)\citenamefont
  {Aldoshin}, \citenamefont {Fel'dman},\ and\ \citenamefont
  {Yurishchev}}]{yuri2}%
  \BibitemOpen
  \bibfield  {author} {\bibinfo {author} {\bibfnamefont {S.}~\bibnamefont
  {Aldoshin}}, \bibinfo {author} {\bibfnamefont {E.}~\bibnamefont {Fel'dman}},
  \ and\ \bibinfo {author} {\bibfnamefont {M.}~\bibnamefont {Yurishchev}},\
  }\href@noop {} {\bibfield  {journal} {\bibinfo  {journal} {Low Temperature
  Physics}\ }\textbf {\bibinfo {volume} {40}},\ \bibinfo {pages} {3} (\bibinfo
  {year} {2014})}\BibitemShut {NoStop}%
\bibitem [{\citenamefont {Ma}\ \emph {et~al.}(2016)\citenamefont {Ma},
  \citenamefont {Yadin}, \citenamefont {Girolami}, \citenamefont {Vedral},\
  and\ \citenamefont {Gu}}]{ma2016converting}%
  \BibitemOpen
  \bibfield  {author} {\bibinfo {author} {\bibfnamefont {J.}~\bibnamefont
  {Ma}}, \bibinfo {author} {\bibfnamefont {B.}~\bibnamefont {Yadin}}, \bibinfo
  {author} {\bibfnamefont {D.}~\bibnamefont {Girolami}}, \bibinfo {author}
  {\bibfnamefont {V.}~\bibnamefont {Vedral}}, \ and\ \bibinfo {author}
  {\bibfnamefont {M.}~\bibnamefont {Gu}},\ }\href@noop {} {\bibfield  {journal}
  {\bibinfo  {journal} {Physical review letters}\ }\textbf {\bibinfo {volume}
  {116}},\ \bibinfo {pages} {160407} (\bibinfo {year} {2016})}\BibitemShut
  {NoStop}%
\bibitem [{\citenamefont {Adesso}\ \emph {et~al.}(2016)\citenamefont {Adesso},
  \citenamefont {Bromley},\ and\ \citenamefont
  {Cianciaruso}}]{adesso2016measures}%
  \BibitemOpen
  \bibfield  {author} {\bibinfo {author} {\bibfnamefont {G.}~\bibnamefont
  {Adesso}}, \bibinfo {author} {\bibfnamefont {T.~R.}\ \bibnamefont {Bromley}},
  \ and\ \bibinfo {author} {\bibfnamefont {M.}~\bibnamefont {Cianciaruso}},\
  }\href@noop {} {\bibfield  {journal} {\bibinfo  {journal} {Journal of Physics
  A: Mathematical and Theoretical}\ }\textbf {\bibinfo {volume} {49}},\
  \bibinfo {pages} {473001} (\bibinfo {year} {2016})}\BibitemShut {NoStop}%
\bibitem [{\citenamefont {Ollivier}\ and\ \citenamefont {Zurek}(2001)}]{zurek}%
  \BibitemOpen
  \bibfield  {author} {\bibinfo {author} {\bibfnamefont {H.}~\bibnamefont
  {Ollivier}}\ and\ \bibinfo {author} {\bibfnamefont {W.~H.}\ \bibnamefont
  {Zurek}},\ }\href@noop {} {\bibfield  {journal} {\bibinfo  {journal}
  {Physical Review Letters}\ }\textbf {\bibinfo {volume} {88}},\ \bibinfo
  {pages} {017901} (\bibinfo {year} {2001})}\BibitemShut {NoStop}%
\bibitem [{\citenamefont {Ciccarello}\ \emph {et~al.}(2014)\citenamefont
  {Ciccarello}, \citenamefont {Tufarelli},\ and\ \citenamefont
  {Giovannetti}}]{ciccarello}%
  \BibitemOpen
  \bibfield  {author} {\bibinfo {author} {\bibfnamefont {F.}~\bibnamefont
  {Ciccarello}}, \bibinfo {author} {\bibfnamefont {T.}~\bibnamefont
  {Tufarelli}}, \ and\ \bibinfo {author} {\bibfnamefont {V.}~\bibnamefont
  {Giovannetti}},\ }\href@noop {} {\bibfield  {journal} {\bibinfo  {journal}
  {New Journal of Physics}\ }\textbf {\bibinfo {volume} {16}},\ \bibinfo
  {pages} {013038} (\bibinfo {year} {2014})}\BibitemShut {NoStop}%
\bibitem [{\citenamefont {Paula}\ \emph {et~al.}(2013)\citenamefont {Paula},
  \citenamefont {de~Oliveira},\ and\ \citenamefont {Sarandy}}]{sarandy2}%
  \BibitemOpen
  \bibfield  {author} {\bibinfo {author} {\bibfnamefont {F.~M.}\ \bibnamefont
  {Paula}}, \bibinfo {author} {\bibfnamefont {T.~R.}\ \bibnamefont
  {de~Oliveira}}, \ and\ \bibinfo {author} {\bibfnamefont {M.~S.}\ \bibnamefont
  {Sarandy}},\ }\href@noop {} {\bibfield  {journal} {\bibinfo  {journal}
  {Physical Review A}\ }\textbf {\bibinfo {volume} {87}},\ \bibinfo {pages}
  {064101} (\bibinfo {year} {2013})}\BibitemShut {NoStop}%
\bibitem [{\citenamefont {Montealegre}\ \emph {et~al.}(2013)\citenamefont
  {Montealegre}, \citenamefont {Paula}, \citenamefont {Saguia},\ and\
  \citenamefont {Sarandy}}]{sarandy3}%
  \BibitemOpen
  \bibfield  {author} {\bibinfo {author} {\bibfnamefont {J.~D.}\ \bibnamefont
  {Montealegre}}, \bibinfo {author} {\bibfnamefont {F.~M.}\ \bibnamefont
  {Paula}}, \bibinfo {author} {\bibfnamefont {A.}~\bibnamefont {Saguia}}, \
  and\ \bibinfo {author} {\bibfnamefont {M.~S.}\ \bibnamefont {Sarandy}},\
  }\href@noop {} {\bibfield  {journal} {\bibinfo  {journal} {Physical Review
  A}\ }\textbf {\bibinfo {volume} {87}},\ \bibinfo {pages} {042115} (\bibinfo
  {year} {2013})}\BibitemShut {NoStop}%
\bibitem [{\citenamefont {Hohenberg}\ and\ \citenamefont
  {Kohn}(1964)}]{hohenberg1964inhomogeneous}%
  \BibitemOpen
  \bibfield  {author} {\bibinfo {author} {\bibfnamefont {P.}~\bibnamefont
  {Hohenberg}}\ and\ \bibinfo {author} {\bibfnamefont {W.}~\bibnamefont
  {Kohn}},\ }\href@noop {} {\bibfield  {journal} {\bibinfo  {journal} {Physical
  review}\ }\textbf {\bibinfo {volume} {136}},\ \bibinfo {pages} {B864}
  (\bibinfo {year} {1964})}\BibitemShut {NoStop}%
\bibitem [{\citenamefont {Maziero}\ \emph {et~al.}(2010)\citenamefont
  {Maziero}, \citenamefont {Guzman}, \citenamefont {C{\'e}leri}, \citenamefont
  {Sarandy},\ and\ \citenamefont {Serra}}]{maziero2010quantum}%
  \BibitemOpen
  \bibfield  {author} {\bibinfo {author} {\bibfnamefont {J.}~\bibnamefont
  {Maziero}}, \bibinfo {author} {\bibfnamefont {H.~C.}\ \bibnamefont {Guzman}},
  \bibinfo {author} {\bibfnamefont {L.~C.}\ \bibnamefont {C{\'e}leri}},
  \bibinfo {author} {\bibfnamefont {M.~S.}\ \bibnamefont {Sarandy}}, \ and\
  \bibinfo {author} {\bibfnamefont {R.~M.}\ \bibnamefont {Serra}},\ }\href@noop
  {} {\bibfield  {journal} {\bibinfo  {journal} {Physical Review A}\ }\textbf
  {\bibinfo {volume} {82}},\ \bibinfo {pages} {012106} (\bibinfo {year}
  {2010})}\BibitemShut {NoStop}%
\bibitem [{\citenamefont {Arnesen}\ \emph {et~al.}(2001)\citenamefont
  {Arnesen}, \citenamefont {Bose},\ and\ \citenamefont
  {Vedral}}]{arnesen2001natural}%
  \BibitemOpen
  \bibfield  {author} {\bibinfo {author} {\bibfnamefont {M.~C.}\ \bibnamefont
  {Arnesen}}, \bibinfo {author} {\bibfnamefont {S.}~\bibnamefont {Bose}}, \
  and\ \bibinfo {author} {\bibfnamefont {V.}~\bibnamefont {Vedral}},\
  }\href@noop {} {\bibfield  {journal} {\bibinfo  {journal} {Physical Review
  Letters}\ }\textbf {\bibinfo {volume} {87}},\ \bibinfo {pages} {017901}
  (\bibinfo {year} {2001})}\BibitemShut {NoStop}%
\bibitem [{\citenamefont {Asoudeh}\ and\ \citenamefont
  {Karimipour}(2006)}]{asoudeh2006thermal}%
  \BibitemOpen
  \bibfield  {author} {\bibinfo {author} {\bibfnamefont {M.}~\bibnamefont
  {Asoudeh}}\ and\ \bibinfo {author} {\bibfnamefont {V.}~\bibnamefont
  {Karimipour}},\ }\href@noop {} {\bibfield  {journal} {\bibinfo  {journal}
  {Physical Review A}\ }\textbf {\bibinfo {volume} {73}},\ \bibinfo {pages}
  {062109} (\bibinfo {year} {2006})}\BibitemShut {NoStop}%
\end{thebibliography}
\end{document}